\documentstyle[11pt,AATS,epsf]{article}	

\newcommand{\etal}{{\em et al.}}
\newcommand{\ie}{{\em i.e.,}}
\newcommand{\eg}{{\em e.g.,}}
\def\gtrsim{\mathrel{\hbox{\rlap{\hbox{\lower4pt\hbox{$\sim$}}}\hbox{$>$}}}}
\begin{document}	

\title{The Oldest Stellar Populations at z $\sim$ 1.5} 

\author{Alan Stockton}
\affil{Institute for Astronomy, University of Hawaii}


\begin{abstract}
There are at least three reasons for being interested in galaxies at
high redshifts that formed most of their stars quite quickly early in
the history of the Universe:  (1) the ages of their stellar populations
can potentially place interesting constraints on cosmological parameters
and on the epoch of the earliest major episodes of star formation, (2)
their morphologies may provide important clues to the history and mechanisms
of spheroid formation, and (3) they are likely to identify the regions of
highest overdensity at a given redshift.

We describe a systematic search for galaxies at $z\sim1.5$ having essentially
pure old stellar populations, with little or no recent star formation.  Our
approach is to apply a ``photometric sieve'' to the fields of quasars near this
redshift, looking for companion objects with the expected spectral-energy
distributions.  Follow-up observations on two of the fields having candidates
discovered by this technique are described.
\end{abstract}


\section{Introduction}

One of the uncertainties in our picture of galaxy formation
in the early Universe is in understanding how
elliptical galaxies and the massive bulges of early-type spirals developed.
The standard cold-dark-matter scenario implies bottom-up formation---small-mass
systems form first, possibly as objects something like the
ubiquitous star-forming dwarf galaxies found at lower redshifts; then these
objects merge to form larger entities, including, in the denser regions,
large spheroidal systems.  However, there are at least two nagging worries
regarding this scenario.
Firstly, there is a correlation between color (\ie\
metallicity) and luminosity in elliptical galaxies (Bower, Lucey, \& Ellis 1992;
Ellis \etal\ 1996).  As Peacock (1999) has observed, ``It seems as if
the stars in ellipticals were formed at a time when the depth of the
potential well that they would eventually inhabit was already determined.''
Secondly, the very tight correlation found between stellar velocity
dispersion in bulges and black-hole mass (Ferrarese \& Merritt 2000; Gebhardt
\etal\ 2000) seems to demonstrate an intimate connection between the formation
of spheroids and the formation of supermassive black holes at their
centers.  It is certainly not obvious that this correlation could be
produced from successive mergers of small building blocks, since 
supermassive black holes do not seem to be associated with pure disks
or irregulars.  In fact, both of these observations would appear to
fit more comfortably with monolithic or quasi-monolithic collapse pictures
of spheroidal formation; however, this statement is more a reflection of
our current uncertainty than an endorsement of such models.  What is
clear is that direct observational constraints on formation mechanisms,
environments, and formation epoch for spheroidals are necessary in
attempting to sort out these difficulties.

\section{The Earliest Major Episodes of Star Formation and Constraints on
Cosmological Parameters}

Figure 1 ({\it left}) shows the $K'$ magnitude of $L^*$ elliptical galaxies 
as a function
of redshift, assuming only passive evolution of a solar metallicity stellar
population formed essentially instantaneously at cosmic epochs of either 0.5 or
1.0 Gyr.  At $z=1.5$, such galaxies would have $K'\sim19.5$, so they are
quite easily detectable.  Both we and others (\eg\ Dunlop 2000 and references
therein) are finding galaxies about 1 mag brighter than this, presumably on
the high-luminosity tail of the luminosity function and indicating that
the stellar content of some early-type galaxies is essentially fully in
place at very high redshifts.

If we could determine precise ages for old populations at high redshifts,
we could potentially place interesting constraints on cosmological 
parameters as well as on formation epochs.  This possibility is shown in 
Fig.~1 ({\it right}):  if one could demonstrate
an age of $\ge4$ Gyr at $z=1.5$, with currently reasonable values of
$h_0$ ($=H_0/100$) and $\Omega_m$ all open models would be eliminated,
and even $\Lambda$-dominated models with $z_f\sim10$ are only barely 
consistent.
\begin{figure}[!h]
\plottwo{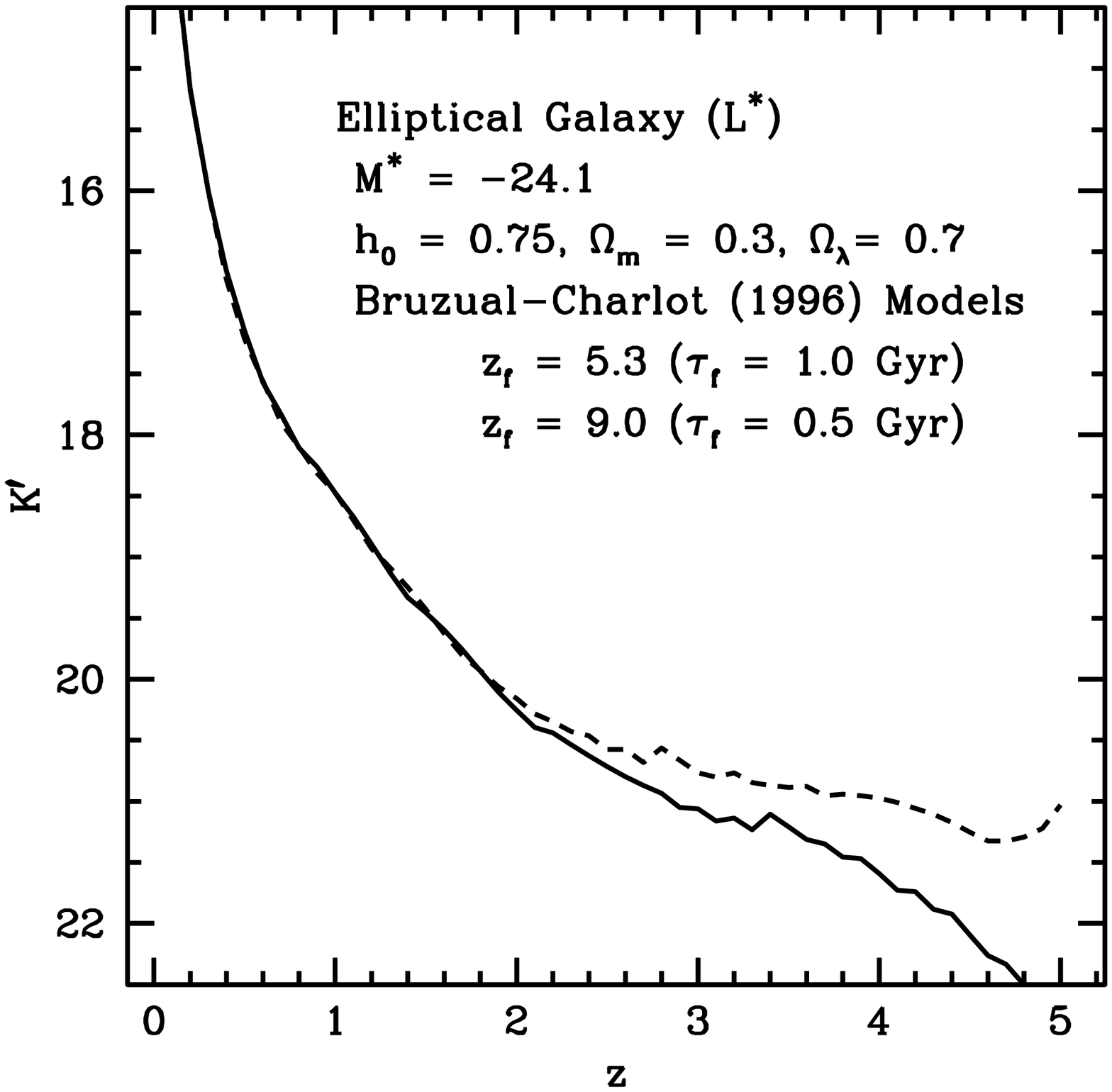}{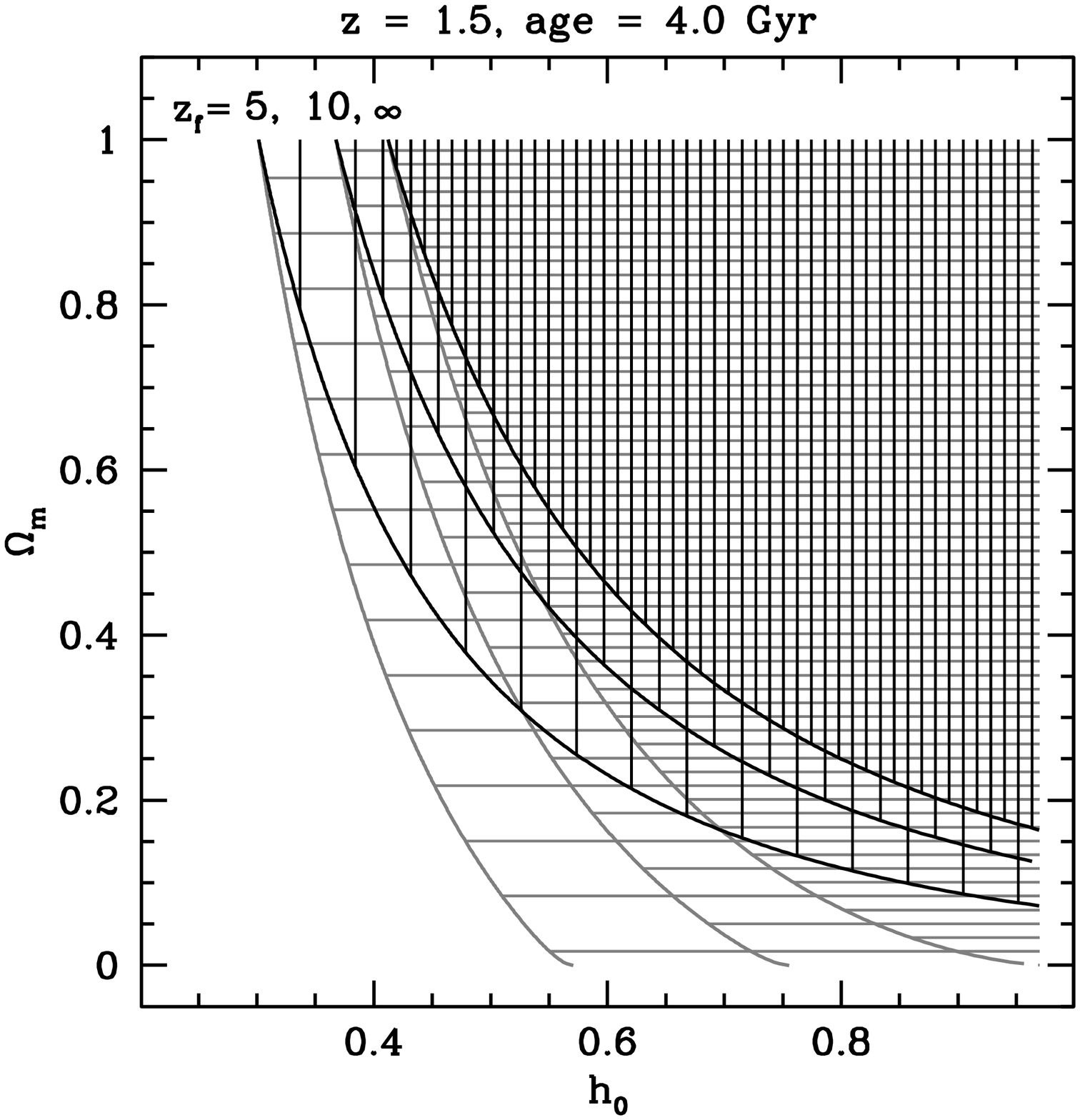}
\caption{({\it left panel})---$K'$ magnitude for a passively evolving $L^*$
elliptical galaxy, assuming an instantaneous burst with formation redshifts
of 5.3 (dotted line) or 9.0 (solid line).  ({\it right panel})---Constraints on
the matter density parameter, $\Omega_m$, and the Hubble parameter, $h_0$
($=H_0/100$) assuming the confirmation of a galaxy at $z=1.5$ with a
4.0-Gyr-old stellar population.  The hatched area to the right of each curve
shows excluded regions, assuming star formation redshift $z_f$ of 5, 10, and
$\infty$, as indicated at the top.  Gray curves are for open models 
($\Omega_{total}=\Omega_m$); black curves are for flat models
($\Omega_m + \Omega_{\Lambda} = 1$).  Even the latter require $z_f > 10$ to
be consistent with $\Omega_m\sim0.3$ and $h_0\sim0.7$.}
\end{figure}

\section{Identifying Old Galaxies at High Redshifts}

There have been three main approaches to identifying old galaxies at high
redshifts:  (1) looking for very red objects among weak radio sources
(Dunlop \etal\ 1996, 2000; Spinrad \etal\ 1997), (2) wide-field multicolor
photometric surveys (Thompson \etal\ 1999; Daddi \etal\ 2000), and 
(3) identifying red objects in
radio source fields (this paper; see also Cimatti \etal\ 1997).  We have 
been examining fields of radio-loud QSOs with $1.4<z<1.78$.
We use a ``photometric sieve'' approach, which gives us
high observing efficiency and clearly distinguishes objects with
old stellar populations and little reddening from heavily reddened objects.
Using the NASA Infrared Telescope Facility, we first image the fields
of interest in the $K'$ band, looking for objects with $18\le K'\le19.5$
within a 30\arcsec\ radius of the quasar.  For fields with such objects,
we then obtain $J$-band imaging, looking for objects with $J\!-\!K'\sim2$.
At this point we have eliminated typically 80\% of our original fields; for
the remainder, we must now obtain CCD imaging on the short side of the
4000 \AA\ break, which occurs between the $I$ and $J$ bands for this
redshift range.  We usually try to obtain at least $R$ and $I$ photometry,
although we have used a variety of standard and non-standard bands.
Recently, a similar approach has been proposed by Pozzetti \& Mannucci (2000).
Of the 208 fields in our sample, we have at least some observations for
about 60\%; we have eliminated 74 fields as having no further interest, and
we have 7 fields with quite firm old-galaxy candidates.
It is significant that 5 of these 7 have more than one good candidate,
including one field with 3.  Spectral-energy distributions for some of these
objects are shown in Fig.~2.
\begin{figure}[!h]
\plotfiddle{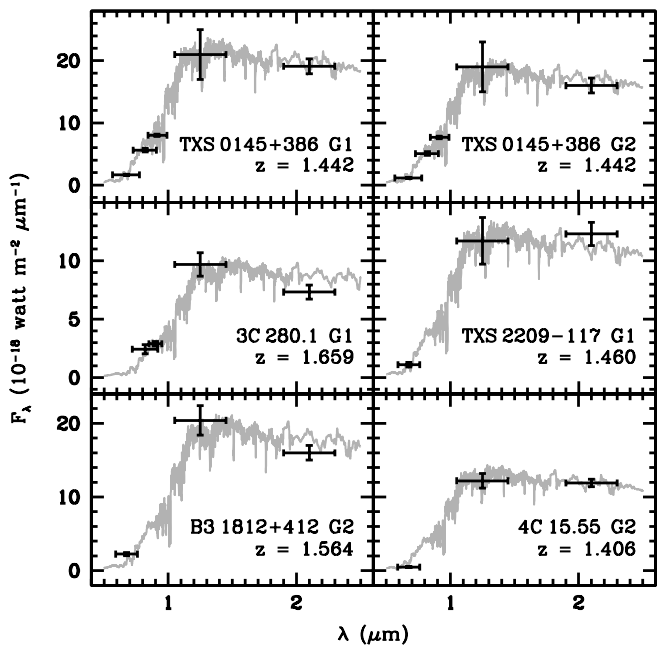}{81mm}{0}{45}{45}{-140}{-75}
\caption{Examples of old galaxies found in fields of quasars at $z\sim1.5$.
Vertical bars show $1\sigma$ photometric errors; horizontal bars show filter
FWHM.  Gray traces are 4 Gyr Bruzual-Charlot (1996) models.}
\end{figure}

\section{An Example: the Field of TXS\,0145+386}
\begin{figure}[!b]
\plotone{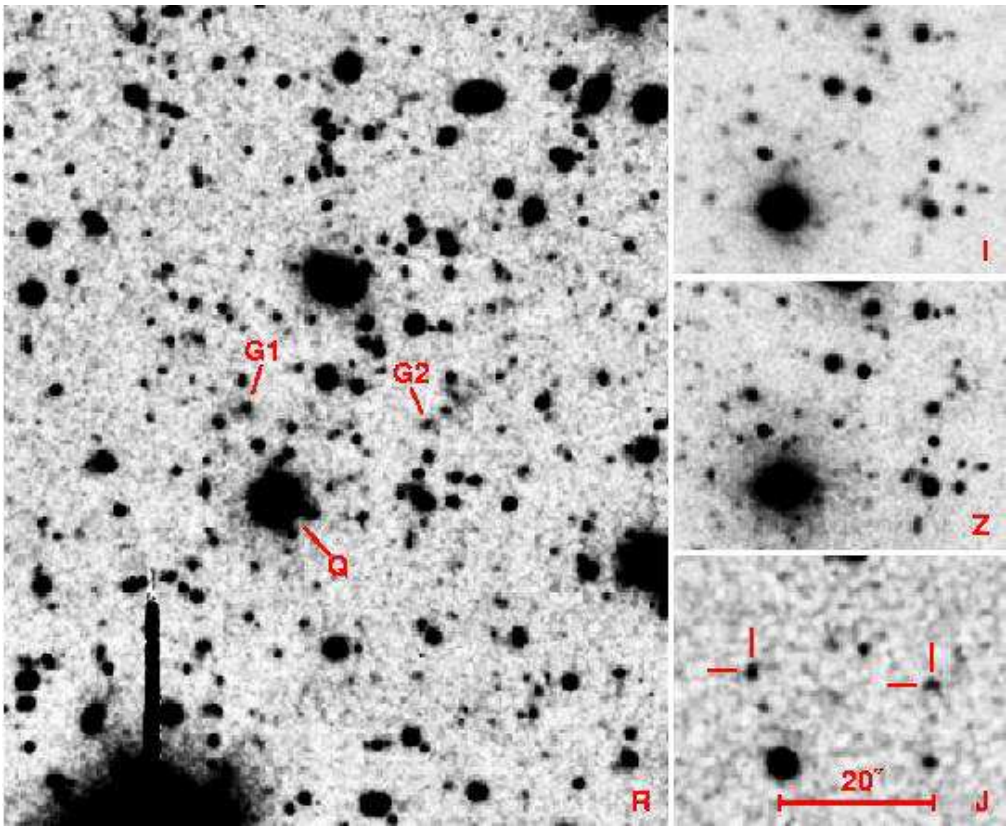}
\caption{Images of the field of TXS$\,$0145+386 in $R$, $I$,
$Z$, and $J$ bands.  The $R$ and $Z$ images were obtained with LRIS on Keck II,
and the $I$ and $J$ images were obtained with the UH 88-inch telescope.  The
quasar (Q) and the two red galaxies, G1 and G2 are marked.  Note the evidence
for a cluster of faint objects in the vicinity of the quasar and red galaxies.
\label{txs0145img}}
\end{figure}

Figure \ref{txs0145img}
shows the field of the $z=1.446$ quasar TXS\,0145+386, in bands ranging from
$R$ to $J$.  Two EROs, G1 and G2, are marked, and their
spectral-energy distributions are shown at the top of Fig.~2.
We have obtained a spectrum of G1, unfortunately through cirrus, so the S/N
is not sufficient to measure age-diagnostic spectral features. However, it does
confirm a redshift of 1.4533 from a weak but broad [\ion{O}{2}] $\lambda3727$ 
line, which possibly indicates the presence of a hidden active nucleus.

\begin{figure}[!t]
\plotone{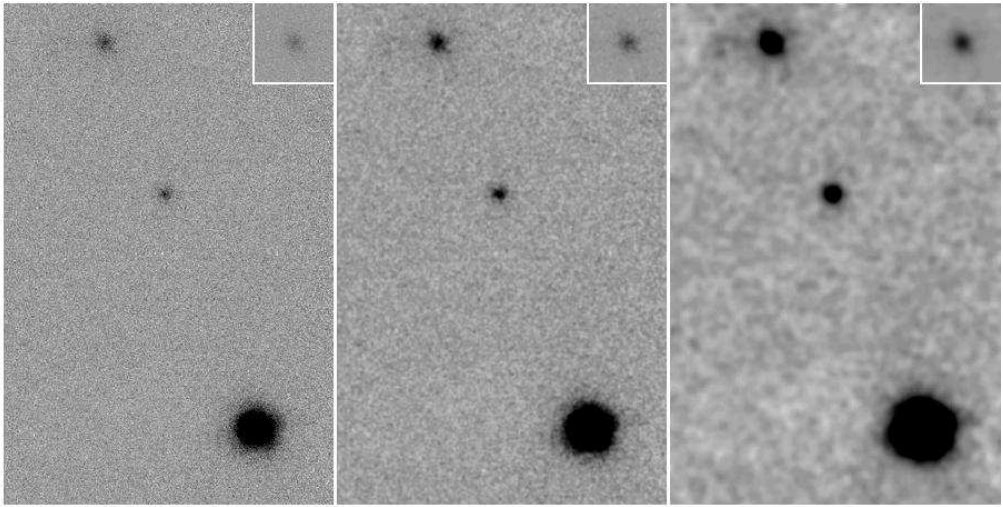}
\caption{Adaptive optics imaging of the field of TXS\,0145+386 at $K'$,
obtained with PUEO on the Canada-France-Hawaii telescope.  The left panel 
shows the unsmoothed image, with
FWHM = 0\farcs16.  The candidate old galaxy G1 is at the top, the nearly stellar
object near the center is a compact galaxy with $z=0.7883$, and the quasar
itself is the bright object at the bottom.  The center and right panels show the
same image, smoothed with a Gaussian with $\sigma=1$ and $\sigma=3$ pixels,
respectively.  Insets show G1 at lower contrast.  The structure
seen to the east of G1 in the right panel appears in both nights' data. 
Panels are $10\arcsec\times15\arcsec$.
\label{txs0145ao}}
\end{figure}

Figure \ref{txs0145ao} shows adaptive optics (AO) imaging of G1.
With images having a FWHM of 0\farcs16, the galaxy is
seen to have a generally symmetric elliptical profile, 
with some faint
irregular structure extending to the east.  The other candidate (G2) is too
far from the guide star for AO imaging, but its morphology on other images
suggests that its structure is less regular (see Fig.~\ref{txs0145img}), in
spite of evidence that its stellar population is as old as that of G1
(Fig.~2).  Candidates in other fields also seem to show a
variety of morphologies.  The implication seems to be that, even with apparent 
stellar
ages of 3--5 Gyrs, these galaxies may not all be completely relaxed systems
in their outer parts, and we may be able actually to observe some aspects
of the final stages of bulge formation.

\begin{figure}[!t]
\plotfiddle{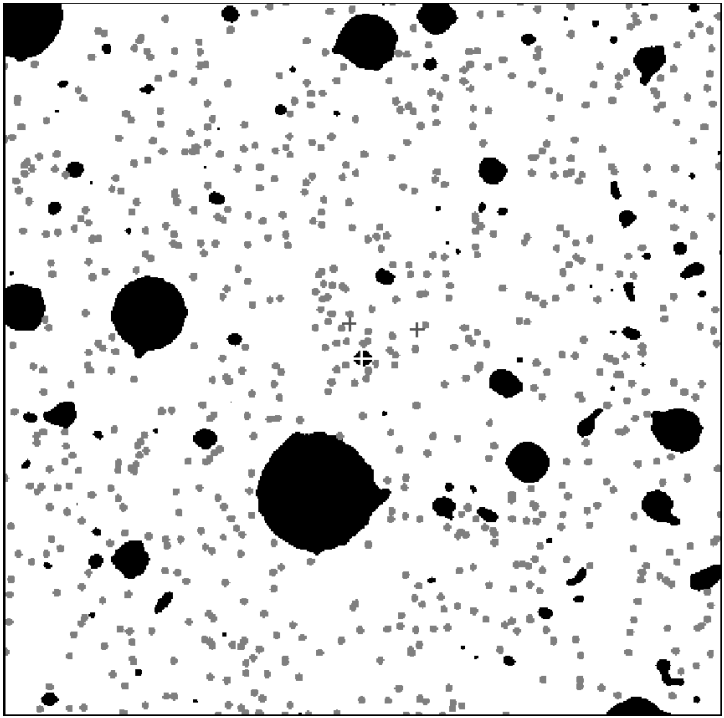}{87mm}{0}{50}{50}{-154}{-67}
\caption{Clustering of galaxies with $24.5<Z_{AB}<26$ (gray dots) in the
field of the quasar TXS\,0145+386 (white cross on black object at center).  The two red galaxies
G1 and G2 are marked with gray crosses.  Regions in black are obscured by bright
stars or galaxies.  The region shown is 4\farcm3 on a side.\label{txs0145cl}}
\end{figure}

There is an apparent
concentration of very faint galaxies in the region around the quasar,
G1, and G2, as seen in Fig.~\ref{txs0145img}.  Figure~\ref{txs0145cl}
gives another view of this clustering.  It is often
suggested that powerful radio sources can be used as markers to locate
regions of high density at high redshifts.
However, while strong radio
sources are undoubtedly statistically in regions of higher density than
is the average galaxy, there appears to be a large dispersion in the
densities of radio source environments.  Some seem to be in fairly
rich clusters (Dickinson, Dey, \& Spinrad 1995; Chapman, McCarthy, \& Persson 2000), but 
some seem
not to be (\eg\ Stockton \& Ridgway 1997).  It may well be that the presence of
nearly fully formed galaxies comprising old stellar populations is a more
reliable indicator of a rich cluster:  under most plausible formation
scenarios, processes of galaxy evolution will proceed more rapidly in strongly
overdense regions.  The identification of such galaxies in radio source fields 
may be one of the best ways of finding rich clusters at redshifts
beyond the practical range of current wide-area X-ray surveys.

\section{Determining Ages of Stellar Populations}
\begin{figure}[!bt]
\plotfiddle{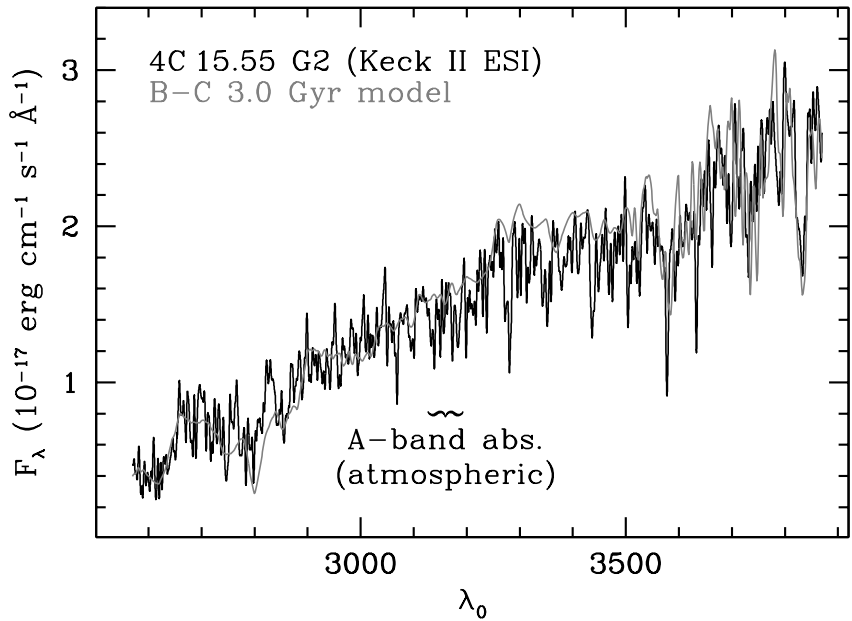}{80mm}{0}{60}{60}{-190}{-140}
\caption{The spectrum of 4C15.55 G2 ($R=26.6$), obtained with ESI on Keck II.
The gray trace shows a Bruzual \& Charlot (1996) 3 Gyr spectral synthesis
model for comparison.}\label{4c15.55spec}
\end{figure}

The main difficulty in attempting to use old galaxies at
moderately high redshifts to constrain cosmological parameters is
in establishing a robust age for the stellar population.
Both we (Stockton, Kellogg, \& Ridgway 1995) and Dunlop \etal\ (1996;
see also Spinrad \etal\ 1997) have attempted to use spectroscopic age
diagnostics.  The spectral features of interest
fall in the rest-frame near-UV (2600--3200 \AA) and potentially give good
age discriminants over the age range $\sim1$--5 Gyr.
Figure \ref{4c15.55spec} shows our preliminary reduction of 
a spectrum of one of two EROs in the field of the $z=1.406$ quasar 4C15.55, 
which fairly closely matches a 3-Gyr-old model.

The major uncertainty in the results so far is in the reliability of the
spectral synthesis models (\eg\ Bruzual \& Magris 1997; Dunlop 1998, 2000;
Heap \etal\ 1998; Yi \etal\ 2000; Nolan \etal\ 2001).  However, as Dunlop
(2000) has emphasized, much of this disagreement is due to the inclusion of
broad-band colors in determining the ages; from the spectroscopic age
diagnostics alone, the age dispersion is much smaller. In addition to being
sensitive to reddening, ages from colors are especially
dependent on getting rather uncertain late stages of stellar evolution right.
On the other hand, while ages derived from restframe near-UV absorption
features require more observing time, they have important advantages:
(1) they are potentially capable of higher precision and depend essentially
only on the turnoff age of main-sequence F stars, where the models are on
much firmer ground;
(2) they are almost totally insensitive to the IMF of the stellar
population; and
(3) they are also insensitive to reddening.
The most serious worry has been the age---metallicity degeneracy (\eg\
Worthy 1994), which can affect both colors and spectral features.  However,
recent work by Nolan \etal\ (2001) indicates that, with sufficiently good
data, it is possible to break this degeneracy from the rest-frame near-UV
spectroscopy alone.
Thus, while it is still quite reasonable to {\it select} candidates on the
basis of colors, it is essential to obtain spectroscopy in the rest-frame
near-UV in order to determine robust lower limits to the age of the
stellar population.  In the meantime, models of increasing sophistication
are being developed by several groups, and such refinements as incorporating
$\alpha$-enhanced stellar atmospheres and evolutionary tracks are likely
to be available soon.

\section{Higher Redshifts}
In parallel with completing our survey for old galaxies in the fields of
quasars with $z\sim1.5$, we are extending the search to higher redshifts.
Now that it seems clearly established that there are galaxies at 
$z\sim1.5$ with ages of $\gtrsim3$ Gyr, we are seeking to identify
precursors to these objects at $z\sim2.5$ with CISCO on the 8.2 m Subaru
telescope.  At this redshift, the oldest galaxies should be $\sim1.5$ Gyr 
younger than at $z\sim1.5$.





\acknowledgments
I wish to acknowledge and thank my collaborators in these projects, Gabriela 
Canalizo, Susan Ridgway, Toshinori Maihara, and Bill Vacca.  This work has 
been partially supported by NSF Grant AST95-29078.
It has also made use of the NASA/IPAC Extragalactic Database (NED), which is
operated by the Jet Propulsion Laboratory, California Institute of Technology,
under constract with the National Aeronautics and Space Administration.
The observations described in this paper were obtained with the NASA
Infrared Telescope Facility, the University of Hawaii 88-inch Telescope,
the Canada-France-Hawaii Telescope, and the Keck Observatories, and I wish
to express my gratitude to
those who have provided these facilities and those who maintain and operate
them.

\end{document}